\documentclass[12pt]{article}
\usepackage{latexsym,amssymb,amsbsy}
\usepackage{epsfig}
\newcommand {\il} {\int\limits}

\newcommand {\sh} {\mathop{\rm sh}\nolimits}

\newcommand {\tr} {\mathop{\rm tr}\nolimits}

\newcommand {\e}{\mathop{\rm e}\nolimits}

\newcommand {\be}    {\begin{equation}}
\newcommand {\ee}    {\end{equation}}
\newcommand {\eeq}    {\end{equation}}
\newcommand {\ba}    {\begin{array}}
\newcommand {\ea}    {\end{array}}
\newcommand {\bea}   {\begin{eqnarray}}
\newcommand {\eea}   {\end{eqnarray}}
\newcommand {\bp}    {\unitlength1mm\begin{picture}}
\newcommand {\ep}    {\end{picture}}
\newcommand {\bmp}   {\begin{minipage}}
\newcommand {\emp}   {\end{minipage}}
\newcommand {\bc}    {\begin{center}}
\newcommand {\ec}    {\end{center}}
\newcommand {\bt}    {\begin{tabular}}
\newcommand {\et}    {\end{tabular}}

\newcommand {\lt}    {\left}
\newcommand {\rt}    {\right}
\newcommand {\ds}    {\displaystyle}

\newcommand {\fr} {\displaystyle\frac}

\renewcommand\vec[1]{{\bf#1}}
\begin{document}
\begin{titlepage}

\title{\bf {Vacuum effects in  electrodynamics and in Yang-Mills
    theory in (2+1) dimensions}}
\date{}
\author{  V.~Ch.~Zhukovsky$^1$, A.~S.~Razumovsky$^1$, and
   K.~V.~Zhukovsky$^2$\\
\small {Faculty of Physics, Moscow State University, 119899 Moscow, Russia}\\
\small {$^1$ Department of
Theoretical Physics }\\
\small {$^2$  Department of Optics
   and Spectroscopy} \\
\small{E-mail: th180@phys.msu.su}}
\unitlength=1pt
\maketitle
\begin{abstract}
Vacuum effects in (2+1)-dimensional
quantum electrodynamics (QED) with the topological Chern-Simons
term are considered. The photon polarization operator  is  studied and
the decay rate for the electron-positron photoproduction $\gamma \to
e^{+}\,\,e^{ - }$  is presented as a function of the photon energy
and external field strength. The radiatively induced electron mass
shift in an external magnetic field is investigated  both taking the
topological Chern-Simons term into account and ignoring
it. Moreover, the electron self-energy in topologically massive
(2+1)-QED at finite temperature and density is studied.
Finally, the parity breaking part of the action in the framework of the
$SU(2)\times U(1)$ gauge field model at finite temperature is considered.
The  massive fermion contribution to the one-loop effective action in
the background of the superposition of an abelian and
a non-abelian gauge fields
leading to parity breaking in the finite temperature (2+1)-quantum
field theory  is discussed.
\end{abstract}

\end{titlepage}

\renewcommand{\thefootnote}{\arabic{footnote}}
\setcounter{footnote}{0}
\setcounter{page}{1}

\tableofcontents
\section{Introduction}

Investigations of quantum  field theory models in
low-dimensional spaces started with a number of discoveries made
in late 1970's and in early 1980's.  In 1979, studies of  linear
polymers were reported in ~\cite{Su}, where it was stated that
the  main
characteristics of the continuum model of polymer chains coincide with
those of the already known one-dimensional models of
quantum fields. Thus, the low-dimensional models proved to be very
useful instruments for studying quasi-one-dimensional and
quasi-two-dimensional media.  Moreover, starting from the
discovery of the integer quantum   Hall effect in 1980 by
von Klitzing and his collaborators ~\cite{Hall_eff}, these models
in the $(2 + 1)$-dimensional space-time have become especially
popular. Recently, a close connection between certain predictions of the
 low-dimensional quantum field theory  and a number of
unusual phenomena detected experimentally in condensed matter
physics has been revealed.

Odd-dimensional gauge theories have attracted much attention since
1981, when S. Deser, R. Jackiw and S. Templeton \cite{K1}, \cite{K3},
and N. Schonfield 
\cite{K2} demonstrated that, in the three dimensional space-time,
a massive gauge invariant field theory can be constructed by
adding (or obtaining due to fermion fluctuations) a topological
Chern-Simons (CS) term $S_{CS}$ to the action of matter and fields.

Topologically massive $(2+1)$-dimensional theories demonstrate a
number of unusual properties. For instance, the finite mass of gauge
fields leads to screening of both electric and magnetic fields
\cite{K3}-\cite{K6}, attraction of
like
charges becomes
possible \cite{K5,K7}, and the requirement of invariance with
respect to topologically nontrivial gauge transformations in
non-abelian theories leads to quantization of the parameter $\theta$
that plays the role of the gauge field mass \cite{K3}.

Interesting properties of $(2+1)$-dimensional theories are related to
statistics.
There are examples of $(3+1)$-dimensional systems with
anionic (fractional) statistics that is intermediate between
the Fermi and Bose statistics  \cite{K8}. In the
$(2+1)$-dimensional world, the anionic statistics, however, is
realized in a unique way \cite{K9,K10}.

As mentioned above, the $(2+1)$-dimensional models are not only
formal illustrative examples, but
they also  have practical applications in the solid state physics,
such as, for instance, high temperature superconductivity or the
quantum Hall effect,  which is explained in the framework of the
two-dimensional anionic model \cite{K11}-\cite{K13}.
The hypothesis that an anionic gas can obtain superconducting properties
was proposed in \cite{K14,K15} after quasi-planar structures were
discovered in high-temperature superconductors \cite{K16}, and it was
confirmed by calculations  in \cite{K16,K17}. After that, the
idea of anionic mechanism of high temperature superconductivity has
become quite popular \cite{K18}-\cite{K21}. In this connection, the
study of  radiative effects in $(2+1)$--dimensional QED under various
external conditions
is on the agenda. In particular, the photon polarization operator
and the Debye length in $(2+1)$--QED at finite temperature and
vanishing chemical potential were studied in \cite{K22}, and at
zero temperature and finite chemical potential in \cite{K23}.

Effects of external gauge fields
\cite{K24}-\cite{K29}, as well as of the gravitational field (see, e.g.,
\cite{pronin}), together with the effects of finite temperature  and
nonzero chemical potential, in various
$(2+1)$-dimensional quantum field models attracted much
interest recently.

This, in particular, is related to the fact that many physical
effects can take place only in the presence of an external field.
For instance, the quantum Hall effect is explained basing on the
peculiarities of the energy spectrum of a two-dimensional electron
gas in a strong magnetic field, and an external magnetic field is
responsible for superconductivity in a two-dimensional system. We
also note that, as it was demonstrated in recent studies of
radiative effects in (2+1)-dimensional theories
\cite{K28}-\cite{Kolya}, the CS topological term plays the role of
an IR regulator parameter even in those cases when an external
field is present, which in itself seems to be able to play the
role of an IR regulator. In particular, the one-loop electron mass
operator  and the photon polarization operator in (2+1)-QED in an
external magnetic field at finite temperature and density
\cite{K28}-\cite{K37}, as well as the quark mass operator in
(2+1)-QCD in an external chromomagnetic field \cite{Kolya} were
calculated. It was demonstrated that these quantities are
described by nonanalytic functions of the external field. With
consideration for the CS term, the exact solutions of the
non-abelian YM field equations in (2+1) dimensions were found and
one-loop vacuum corrections to the effective lagrangian of these
fields due to fluctuations of gauge and fermion fields were
calculated \cite{peskov}.  We also note that the problem of vacuum
effects in low dimensional field theories is of special interest
due to the development of non-abelian gauge field theories at high
temperatures and in strong external fields, where the effective
dimensional reduction takes place (see, e.g., \cite{K30} and
references therein).
Recently, various effects in odd-dimensional theories have been
studied with consideration for the influence of external
conditions (see, e.g., investigations of the photon polarization
operator and the electron mass operator in the $(2+1)$-dimensional
quantum electrodynamics in a magnetic field at finite temperature
and density \cite{Kostya,K29,book}).

It should also be mentioned that the problem of dynamical
generation of the Chern-Simons term in certain models was
considered in detail in \cite{Klim1} and~\cite{Klim2}. Note also
the discussion of the role of the finite matter density in the
effect of dynamical generation of the CS term in odd-dimensional
gauge theories (see, e.g.,\cite{tseit,sissak}), which demonstrates
that the scope of the study of topological problems in gauge
theories is still far from being exhausted.

The fundamental property of the gauge field action in the
non-abelian case is its noninvariance with respect to the so
called ''large" (homotopically nontrivial) gauge transformations.
Therefore, to make the path integral, which involves
$\exp{(iS_{CS})}$, invariant under these transformations, the
requirement that the CS coefficient (the topological mass) is
quantized in units of $g^2/(4 \pi)$ should be adopted. However,
the above holds only for the zero temperature quantum fields. When
the temperature is finite, situation becomes quite different. This
relates to the fact that perturbative corrections to the CS term
are non-quantized continuous functions of temperature \cite{2}. As
a result, "large" gauge invariance would be lost at finite
temperature. Nevertheless, as it was demonstrated by S. Deser,
L. Griguolo, and D. Seminara 
\cite{D1}--\cite{D4}, even though the CS term 
itself may violate "large" gauge invariance, there exist other
terms in the effective action, which can compensate for this
violation and make the total effective action gauge invariant
under both small and large gauge transformations.
At first, it was explicitly shown
for the (0+1) dimensional model \cite{3,7}. Then, this analysis
was generalized to (2+1) dimensional fermions interacting
with Abelian \cite{D1,D2,D3,4} and non-Abelian \cite{D1,D4} gauge
backgrounds. There, in the framework of the non-perturbative approach, it was
clearly demonstrated, that even though the perturbative expansion
leads to a non-quantized temperature-dependent CS coefficient, the
full effective action can be made invariant under "large" gauge
transformations (LGT) at any temperature, once the suitable
regularization of the Dirac operator determinant is applied. As a result,
there arises a parity anomaly as a price for the 
restoration of the invariance.


Of special interest are recently proposed various
modifications of the (3+1)-dimensional QED. For instance, a Chern-Simons
like term can be added
to the QED Lagrangian \cite{jackiw} with the coupling of the dual
electromagnetic field tensor to a certain fixed 4-vector, or
an additional Lorentz-noninvariant term can be introduced in the fermionic
Lagrangian \cite{jak,andrianov}.

In the present review, as an illustration of the above
observations, we present some results of calculations of vacuum
effects in (2+1)-dimensional QED. In Section 2, some basic results of the
(2+1)-dimensional QED are formulated. Further, the study of the
photon polarization operator is presented  and the rate of the
photo-production of electron-positron pair $\gamma \to
e^{+}\,\,e^{ - }$ as a function of the photon energy and of the
external field strength is calculated (Section 3). In Section 4,
we investigate the radiatively induced electron mass shift in an
external magnetic field with the topological Chern-Simons term as
well as without it. The electron vacuum  energy in the
topologically massive (2+1)-QED at finite temperature and density
is studied in Section 4.

Section 5 is devoted to the topological effects in (2+1)-dimensional
$SU(2)\times U(1)$-gauge field theory with the superposition of
the Yang-Mills field ($SU(2)$ group) and the Maxwell field ($U(1)$
group). As an example, we have calculated the exact expression for
the parity breaking part of the effective action in this model. In
conclusion, the analysis of the results obtained is presented.

\section{(2+1)-Dimensional Quantum Electrodynamics}

The topologically massive $(2+1)$-dimensional quantum electrodynamics
is described by the following Lagrangian \cite{K3}:
\begin{equation}
{\cal L}=-\frac{1}{4}F_{\mu\nu}F^{\mu\nu}+
\bar\psi\left[\gamma^{\mu}(i\partial_{\mu}-e A_{\mu})-m \right]\psi+\frac{1}{4}\theta \varepsilon^{\mu\nu\alpha}F_{\mu\nu} A_{\alpha},
\label{K1}
\end{equation}
where $F^{\mu\nu}=\partial^{\mu} A^{\nu} -\partial^{\nu} A^{\mu},$ and
matrices $\gamma^{\mu}$ $(\mu=0,1,2),$  coincidig with the Pauli matrices,
satisfy the folloing relations
\begin{equation}
\label{eq2}
\gamma ^{0} = \sigma ^{3},
\quad
\gamma ^{1} = i\sigma ^{1},
\quad
\gamma ^{2} = i\sigma ^{2},
\end{equation}
\begin{equation}
\label{eq3}
\{ \gamma ^{\mu} ,\gamma ^{\nu} \} = 2g^{\mu \nu} ,
\quad
\gamma ^{\mu} \gamma ^{\nu}  = g^{\mu \nu}  - i\varepsilon ^{\mu \nu \lambda
}\gamma _{\lambda}  ,
\quad
g^{\mu \nu}  = {\rm diag}\left( {1, - 1, - 1} \right).
\end{equation}

The last term in (\ref{K1}) is the so called topological Chern-Simons
term with the coefficient $\theta$ as the gauge field mass.

Under the gauge transformations $A_{\mu}  \to A_{\mu}  +
{\displaystyle\frac{{1}}{{e}}}\partial _{\mu}  \alpha $, $\varphi
\to \e^{i\alpha} \varphi $, $\bar {\varphi}  \to \e^{-i\alpha}
\bar {\varphi} $, the Lagrangian is changed by adding the full derivative
${\cal L} \to {\cal L} + \partial _{\rho}  \bigg(
{{\displaystyle\frac{{\theta} }{{4e}}}\varepsilon ^{\rho \mu \nu}
F_{\mu \nu} \alpha} \bigg)$. The Chern-Simons term together with
the fermion mass term lead to breaking P and T symmetries,
while PT and CPT symmetries are conserved.

It should be mentioned that even if no CS term is present initially in
the Lagrangian (\ref{K1}) of the theory with broken P and
T symmetries, this term is generated
by quantum fluctuations of massive fermions. The magnitude of this
terms significantly
depends both on  finite temperature and density
\cite{K31,K32}, and on the external field strength
\cite{K28}-\cite{K37}. In the one-loop approximation, this can be made
clear by calculating the photon polarization operator (PO). First of
all, consider the photon propagator in QED with the topological term. It can be
written in the following form \cite{K3} (in the general Landau gauge):
\begin{equation}
\label{eq6}
D_{\mu \nu}  \left( {k} \right) = - \frac{{i}}{{k^{2} - \theta ^{2} +
i\varepsilon} }\left( {g_{\mu \nu}  - \frac{{k_{\mu}  k_{\nu} } }{{k^{2}}} +
i\theta \frac{{\varepsilon _{\mu \nu \lambda}  k^{\lambda} }}{{k^{2}}}}
\right).
\end{equation}
This expression clearly demonstrates that the mass of the gauge
field is equal to $\theta $.

The one-loop photon PO in (2+1)-QED is determined by the
expression:
\begin{equation}
\label{eq7}
 \Pi_{\mu\nu}(k)=e^2\int
\frac{d^3k}{( 2\pi)^3}{\rm Tr}\bigg[\gamma
_{\mu}S_c\bigg(x'x\bigg)\gamma _{\nu}S_c\bigg({x\,x'}
\bigg)\bigg].
\end{equation}
It should be emphasized that the polarization
operator $\Pi_{\mu\nu} (k)$ in (3+1)-QED is a symmetric tensor, while in the
(2+1)-QED it is represented as a sum of two terms
\begin{equation}
\Pi_{\mu\nu}=\Pi_{\mu\nu}^S + \Pi_{\mu\nu}^A ,
\label{sa}
\end{equation}
one of them $\Pi_{\mu\nu}^S$ being symmetric, and the other
$\Pi_{\mu\nu}^A$ antisymmetric. The so called dynamically induced
(antisymmetric) part of the CS term in the PO has the following
structure:
\begin{equation}
\Pi_{\mu\nu}^{A}(q)=i\varepsilon_{\mu\nu\alpha}q^{\alpha}\Pi^{A}(q^2) ,
\label{K3}
\end{equation}
Here, the quantity $\Pi^{A}(q^2)$ at $q^2=0$ determines the
Chern-Simons mass induced by radiative effects \cite{K3,K32}
\begin{equation}
\theta_{ind}=\lim_{q \to0}\Pi^{A}(q^2).
\label{K2}
\end{equation}
It should be mentioned that higher order terms of the perturbation
theory do not contribute to  $\theta_{ind}$ (see also \cite{K33,K34}).

Calculations of the photon PO as well as of the electron mass
operator require the knowledge of the fermion propagator in the
external magnetic field, which is determined by the following
equation:
\begin{equation}
\label{eq11}
\left[ {\left( {i\partial _{\mu}  - eA_{\mu} }  \right)\gamma ^{\mu}  - m}
\right]G\left( {x,x'} \right) = \delta ^{3}\left( {x - x'} \right),
\end{equation}
where $G(x, x')$ and $S(x, x')$ are connected by the relation
\begin{equation}
\label{eq12}
G\left( {x,x'} \right) = \left[ {\left( {i\partial _{\mu}  - eA_{\mu} }
\right)\gamma ^{\mu}  + m} \right]S\left( {x,x'} \right).
\end{equation}
For a constant magnetic field, given by the potential
\begin{equation}
\label{eq13}
A^{\mu}  = \left( {0,0,x^{1}H} \right),\;F_{12} = - F_{21} = H,
\end{equation}
calculations of the electron propagator are conveniently performed by the
Fock-Schwinger proper-time
method \cite{fsch,sch}, resulting in the following expression \cite{green}:
\begin{equation}
\label{eq14}
{\begin{array}{l}
 { G\left( {x,x'} \right)=
{\displaystyle\frac{\e^{i\pi /4}eH} {\left( {4\pi}
\right)^{3/2}}}\bigg[ {\gamma ^{\mu }\left( {i\,\partial_{\mu_{x}}
- eA_{\mu}  \left( {x} \right)} \right) + m} \bigg]
{{\displaystyle\int\limits_{0}^{\infty}{\frac{{ds}}{{s^{1/2}}}}\exp{\bigg(-ie\int\limits_{x}^{x'}
{d\xi ^{\mu}A_{\mu}\left({\xi}\right)}
}}\bigg)}\times}\\
 {\;} \\
\times\bigg( {I\,\cot(eHs) - i\sigma ^{3}} \bigg)\exp\bigg({ -
{\displaystyle\frac{{i}}{{4}}}\bigg[{\displaystyle\frac{{\Delta
x_{0} ^{2}}}{{s}}} - \Delta x^{2}eH\cot(eHs)\bigg] -
is\left({m^{2}
- i\varepsilon} \right) }\bigg),  \\
 \end{array}}
\end{equation}
where $m$ is the electron mass, $I = \left( {{\begin{array}{*{20}c}
 {1} \hfill & {0} \hfill \\
 {0} \hfill & {1} \hfill \\
\end{array}} } \right)$, $\Delta x_{1} = x_{1} - x'_{1} $, $\Delta x_{2} =
x_{2} - x'_{2} $, $\Delta x_{0} = x_{0} - x'_{0} $, $\Delta x^{2}
= \left( {\Delta x_{1}}  \right)^{2} - \left( {\Delta x_{2}}
\right)^{2}$. The substantial difference between the above
expression and the expression for the electron propagator in (3+1)-QED
is in the preexponential factor in the integral over $s$: it
is $1/s^{1/2}$ instead of $1/s$ in the (3+1)-QED.

The important feature of the fermion spectrum in the external field in
the (2+1)-dimensional QED is its discreteness:
\begin{equation}
p_{0} = \pm \left( {m^{2} +
2\left| {eH} \right|k} \right)^{1/2}, \,\,k=0,1,2,\dots.
\label{discr}
\end{equation}
This spectrum is typical
for fermions in  space-times of reduced dimensions, since  the degree of
freedom that corresponds to motion along the field is missing in this
case, contrary to the (3+1)-dimensional QED. In
that last case, the fermion spectrum is given by the expression that
depends also on the projection $p_3$ of the electron momentum on the
direction of the field vector, taking continuous values, i.e.,
\[p_{0} = \pm \left( {m^{2} + 2\left| {eH} \right|n +
p_3}  \right)^{1/2}, \,\,n=0,1,2,\dots.
\]
One more special feature of the (2+1)-QED should be mentioned:
contrary to (3+1)-QED with the dual field tensor
\begin{equation}
\label{eq15}
\tilde {F}_{\mu \nu}  = {1\over2}\varepsilon _{\mu \nu \lambda \rho}  F^{\lambda \rho
}
\end{equation}
no such tensor can be constructed in the (2+1)-QED, due to the absence
of the 4$^{th}$ rank  tensor like $\varepsilon_ {\mu \nu \lambda
\rho} $.

\section{ Photon polarization operator  in (2+1)-dimensional QED}

\subsection{ Polarization operator and  photon  elastic scattering amplitude
in a constant magnetic field}

\begin{figure}
\begin{center}
\includegraphics[scale=0.8]{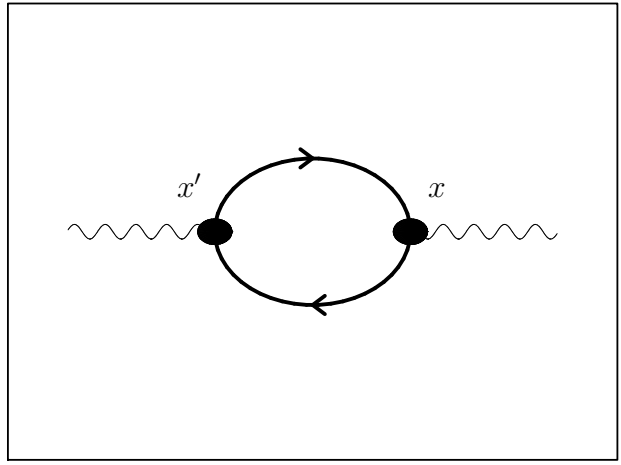}
\begin{picture}(0,0)(0,0)
\put(-170,100){$x'$} \put(-75,100){$x$}
\end{picture}
\caption{\label{PO}Photon polarization operator in the one-loop
approximation.}
\end{center}
\end{figure}

In order to calculate the photon elastic scattering amplitude,
consider the photon polarization operator, written in the one-loop
approximation according to (\ref{eq14}). The equivalent graphical
representation is given in Fig.\ref{PO}
In the constant magnetic field, determined by the potential
(\ref{eq13}), electron propagator (\ref{eq14}) can be transformed
into the following form \cite{K28}:
\begin{eqnarray}
\label{eq17}
 S\left( {x,x'} \right) =\!\!\!\!\!\!\!\!\!\!&&- {\displaystyle\frac{{\e^{ - i\pi /4}h}}{{\left(
{4\pi}  \right)^{3/2}}}}{\displaystyle\int\limits_{0}^{\infty}
{\frac{{ds}}{{\sqrt {s} }}\frac{{\e^{ - ism^{2}}}}{{\sin\left(
{hs} \right)}}}}\exp\bigg( { - i{\displaystyle\frac{{X_{0}
^{2}}}{{4s}}} + i{\displaystyle\frac{{X_{ \bot}  ^{2}h\,\cot\left(
{hs}
\right)}}{{4}}} - iuY{\displaystyle\frac{{h}}{{2}}}} \bigg) \times \nonumber \\
 &&\times\left[ {{\displaystyle\frac{{1}}{{2s}}}\bigg( {\gamma ^{0}T - \frac{{hs}}{{\sin\left( {hs}
\right)}}\left( {\gamma X} \right)_{ \bot}  \e^{ihs\gamma ^{0}}}
\bigg) + m} \right]\e^{ - ihs\gamma ^{0}},
\end{eqnarray}
where $h = eH$ and
\begin{equation}
\label{eq18}
X^{\mu}  = x^{\mu}  - x'^{\mu} ,
\quad
u = x^{1} + x'^{1},
\quad
Y = x^{2} - x'^{2},
\quad
X_{ \bot}  ^{2} = \left( {x^{1} - x'^{1}} \right)^{2} + \left( {x^{2} -
x'^{2}} \right)^{2}.
\end{equation}
Calculation of the photon polarization operator (\ref{eq7}) is
performed similar to the (3+1)-dimensional QED \cite{K38}. In the
(2+1)-dimensional QED,  (\ref{eq7}) depends on the kinetic
momentum operator $\hat {P}_{\mu}=i\partial_{\mu}-ieA_{\mu}$ and
operators  $F^{\mu \nu} \hat {P}_{\nu},\,\ F\;^{\mu \nu} F_{\nu
\lambda}  \hat {P}^{\lambda},$ and thus, it commutes with operator
$\hat {P}_{\mu}  $  in a constant magnetic field. Therefore, the
photon PO turns out to be diagonal in the momentum $k$-space:
\begin{equation}
\label{eq19} \Pi ^{\mu \nu} \left( {k,k'} \right) = \int {\Pi
_{\mu \nu}  \left( {x,x'} \right)\exp\bigg( { - i\left( {kx - kx'}
\right)} \bigg)} dxdx' = \left( {2\pi}  \right)^{3}\delta \left(
{k - k'} \right)P_{\mu \nu}  \left( {k} \right)
\end{equation}
As was already mentioned above, in contrast to (3+1)-QED, where the
polarization operator is symmetrical, it is represented in
(2+1)-QED as a sum of symmetric and antisymmetric terms (\ref{sa}).

Consider first the symmetric part $P_{\mu \nu}  ^{s}\left( {k,H}
\right)$ and expand it over its eigenvectors
$V_\mu^{(i)}$:
\begin{equation}
\label{eq20}
P_{\mu \nu}^{s}\left( {k,H} \right) = \sum\limits_{i = 1}^{3} {\lambda
_{i} \frac{{V_{\mu}  ^{\left( {i} \right)}V_{\nu}  ^{\left( {i}
\right)}}}{{\left| {V^{i}} \right|^{2}}}} \quad ,
\end{equation}
where
\begin{equation}
V_{\mu}  ^{\left( {1} \right)} \equiv l_{\mu}  = \frac{{F_{\mu
\nu}  k^{\nu }}}{{\left( {F_{\lambda \rho}  ^{2}/2}
\right)^{1/2}}}, \quad V_{\mu}  ^{\left( {2} \right)} \equiv
v_{\mu}  = k_{\mu}+\frac{{k^{2}F_{\mu \nu}  l^{\nu}
}}{{l^{2}\left( {F_{\lambda \rho}  ^{2}/2} \right)^{1/2}}}, \quad
V_{\mu}  ^{\left( {3} \right)} = k_{\mu}, \label{2.6}
\end{equation}
$$
\frac{{l_{\mu}  l_{\nu}
}}{{l^{2}}} + \frac{{v_{\mu}  v_{\nu} } }{{v^{2}}} + \frac{{k_{\mu}  k_{\nu
}} }{{k^{2}}} = g_{\mu \nu}.
$$
One of the eigenvalues of $P_{\mu \nu}  ^{s}\left( {k,H} \right)$
vanishes, $\lambda_{3}=0$, since the polarization operator $\Pi_{\mu
\nu} $ is transversal due to its gauge invariance:
\begin{equation}
\label{eq21}
k_{\mu}  \Pi ^{\mu \nu}  = 0
\end{equation}

Since the dual tensor can not be constructed in (2+1) dimensions,
the eigenvector
\begin{equation}
\label{eq22}
V_{\mu}  ^{\left( {4} \right)} = \frac{{\tilde {F}_{\mu \nu}  k^{\nu
}}}{{\left( {F_{\lambda \rho}  ^{2}/2} \right)^{1/2}}}
\end{equation}
 does not exist either, so the number of eigenvectors $V_{\mu}^{(i)}
$ of $P_{\mu \nu}  ^{s}\left( {k,H} \right)$ is also reduced to
three, and only two of the eigenvalues $\lambda _{i}$ are not
vanishing. The symmetric part of the polarization operator can be
expanded as follows:
\begin{equation}
\label{eq23}
P_{\mu \nu}  ^{s}\left( {k,H} \right) = \lambda _{1} \frac{{l_{\mu}  l_{\nu
}} }{{l^{2}}} + \lambda _{2} \frac{{v_{\mu}  v_{\nu} } }{{v^{2}}} = \left(
{\lambda _{1} - \lambda _{2}}  \right)\frac{{l_{\mu}  l_{\nu} } }{{l^{2}}} +
\lambda _{2} \left( {g_{\mu \nu}  - \frac{{k_{\mu}  k_{\nu} } }{{k^{2}}}}
\right)
\end{equation}
where
\begin{equation}
\label{eq24}
l_{\mu}  = \left( {0,k_{2} , - k_{1}}  \right),
\quad
v_{\mu}  = \left( {k_{0} ,\frac{{k_{0} ^{2}}}{{\vec {k}^{2}}}k_{1}
,\frac{{k_{0} ^{2}}}{{\vec {k}^{2}}}k_{2}}  \right).
\end{equation}

Calculating the eigenvalues $\lambda _{1}$, $\lambda _{2}$ from
the eigenvalue equation in the form
\begin{equation}
\label{eq25} \lambda _{1} = \frac{{1}}{{l^{2}}}P_{\mu \nu}
^{s}l^{\mu} l^{\nu}, \quad \lambda _{2} =
\frac{{1}}{{v^{2}}}P_{\mu \nu}  ^{s}v^{\mu} v^{\nu},
\end{equation}
we obtain  for them the following result that takes account of the
external field exactly and allows to write the symmetric part of
the photon polarization operator (\ref{eq23}), (\ref{eq24}):
\begin{eqnarray}
\label{eq26}
 \left( {{\begin{array}{*{20}c}
 {\lambda _{1}}  \hfill \\
 {\lambda _{2}}  \hfill \\
\end{array}} } \right) \!\!\!\!\!\!\!\!&&= - {\displaystyle\frac{{\e^{i\pi /4}e^{2}h}}{{\left( {4\pi}
\right)^{3/2}}}}{\displaystyle\int\limits_{0}^{\infty}  {ds}
\int\limits_{0}^{\infty} {dt\frac{{1}}{{\left( {s + t}
\right)^{1/2}\sin\left( {s + t} \right)}}}}\times \nonumber \\
\nonumber
\\
 &&\times\exp\Bigg( {i\bigg[ {{\displaystyle\frac{{st}}{{s + t}}}k_{0} ^{2} - {\displaystyle\frac{{\sin\left( {hs} \right)\sin\left( {ht}
\right)}}{{h\sin\left( {h( {s + t})} \right)}}}\vec {k}^{2} -
m^{2}\left( {s + t} \right)} \bigg]} \Bigg)\left(
{{\begin{array}{*{20}c}
 {A_{1}}  \hfill \\
 {A_{2}}  \hfill \\
\end{array}} } \right),
\end{eqnarray}
\begin{eqnarray}
\label{eq27}
 A_{1} \!\!\!\!\!\!\!\!\!&&= 2m^{2}\cos\left( {h\left( {t - s} \right)} \right) - 2imk_{0}
\sin\left( {h\left( {s - t} \right)} \right) + \left( { - i +
{\displaystyle\frac{{2stk_{0} ^{2}}}{{s + t}}}}
\right){\displaystyle\frac{{\cos\left( {h\left( {s - t} \right)}
\right)}}{{s + t}}}- \nonumber \\  \nonumber
\\&&-
2{\displaystyle\frac{{\sin\left( {hs} \right)\sin\left( {ht}
\right)}}{{\sin^{2}\left( {h\left( {s + t} \right)} \right)}}}\vec
{k}^{2},
\end{eqnarray}
\begin{eqnarray}
\label{eq28}
 A_{2}\!\!\!\!\!\!\!\!\!&&= A_{1} + 2i{\displaystyle\frac{{\vec {k}^{2}}}{{k^{2}}}}\left[
{{\displaystyle\frac{{h}}{{\sin\left( {h\left( {s + t} \right)}
\right)}}} - {\displaystyle\frac{{\cos\left( {hs}
\right)\cos\left( {ht} \right)}}{{s + t}}}} \right] +
4{\displaystyle\frac{{\vec {k}^{2}}}{{k^{2}}}}m^{2}\sin\left( {hs}
\right)\sin\left( {ht} \right) - \nonumber \\  \nonumber
\\
 &&-4k_{0} ^{2}{\displaystyle\frac{{\vec {k}^{2}}}{{k^{2}}}}\left[ {{\displaystyle\frac{{t\sin\left( {hs}
\right) + s\sin\left( {ht} \right)}}{{2\left( {s + t}
\right)\sin\left( {h\left( {s + t} \right)} \right)}}} -
{\displaystyle\frac{{\sin\left( {hs} \right)\sin\left( {ht}
\right)}}{{\sin^{2}\left( {h\left( {s + t} \right)} \right)}}} -
{\displaystyle\frac{{st}}{{\left( {s + t} \right)^{2}}}}\cos\left(
{hs} \right)\cos\left( {ht} \right)} \right].
\end{eqnarray}
The antisymmetric part of the PO, calculated with account for
formulas (\ref{eq2}), (\ref{eq7}), (\ref{K3}),
(\ref{eq17})-(\ref{eq19}), becomes
\begin{eqnarray}
\label{eq29}
 P^{\mu \nu} _{a} \left( {k,H} \right) \!\!\!\!\!\!\!&&= 2im\e^{i\pi
/4}{\displaystyle\frac{{e^{2}H}}{{\left( {4\pi}
\right)^{3/2}}}}\varepsilon ^{\mu \nu \lambda} k_{\lambda}
\int\limits_{0}^{\infty}  {ds}
{\displaystyle\int\limits_{0}^{\infty } {dt} \frac{{\cos\left(
{h\left( {s - t} \right)} \right)}}{{\sqrt {s + t} \sin\left(
{h\left( {s + t} \right)} \right)}}}\times \nonumber \\ \nonumber
\\
 &&\times\exp\bigg( {i\bigg[ {{\displaystyle\frac{{st}}{{s + t}}}k_{0} ^{2} - \vec
{k}^{2}{\displaystyle\frac{{\sin\left( {hs} \right)\sin\left( {ht}
\right)}}{{h\sin\left( {h\left( {s + t} \right)} \right)}}} -
m^{2}\left( {s + t} \right)}\bigg]} \bigg)
\end{eqnarray}

This part of the PO determines the induced Chern-Simons mass in
(2+1)-QED, according to (\ref{K3}), (\ref{K2}). Thus, the
Chern-Simons term can be induced not only by the finite
temperature and density ($T \ne 0, \mu  \ne 0$) \cite{K31,K32},
but also by the effects of the external field. This is the
important conclusion in considering the topologically massive QED,
especially in the case, when the initial Lagrangian does not
contain a term with the Chern-Simons mass.

The polarization operator, calculated on the mass shell, $k^2=0$,
determines the   photon elastic   scattering amplitude:
\begin{equation}
\label{eq30} T = \frac{{1}}{{2\omega} }e_{\mu}  P^{\mu \nu} _{\rm
reg} e_{\nu}  ,
\end{equation}
where $\omega = k + 0 = \left| {\vec {k}} \right|$ is the photon
energy, $e_\mu $ is the photon polarization vector. The photon PO
is renormalized in the standard way:
\begin{equation}
\label{eq31} P^{\mu \nu} _{\rm reg} \left( {k,H} \right) = P_{\mu
\nu}  \left( {k,H} \right) - P_{\mu \nu}  \left( {k,H = 0} \right)
+ P_{\mu \nu}  \left( {k} \right),
\end{equation}
where $P_{\mu \nu}(k)$ is the renormalized  polarization operator
in the absence of the field. On the mass shell, $m=0$, the eigenvector
$v_\mu^{(2)}=k_\mu $ and hence the PO in (2+1)-QED is determined
by a single vector of linear polarization. It can be represented
in the following form:
\begin{equation}
\label{eq32}
e_{\mu}  = \frac{{l_{\mu} } }{{\left( { - l^{2}} \right)^{1/2}}} =
\frac{{1}}{{\left| {\vec {k}} \right|}}\left( {0,k_{2} , - k_{1}}  \right),
\quad
l_{\mu}  = \frac{{F_{\mu \nu}  k^{\nu} }}{{\left( {F_{\lambda \rho}  ^{2}}
\right)^{1/2}}}.
\end{equation}
 Finally, we  obtain the photon elastic scattering amplitude:
\begin{eqnarray}
\label{eq33}
 T \!\!\!\!\!\!\!\!&&= \e^{i\pi /4}{\displaystyle\frac{{e^{2}m}}{{\left( {4\pi}  \right)^{3/2}\omega} }}{\displaystyle\sqrt
{2\frac{{H_{0}} }{{H}}}} {\displaystyle\int\limits_{ - 1}^{1} {dv}
\int\limits_{0}^{\infty } {\sqrt {\rho}  d\rho}}  \exp\left( { -
2i\rho {\displaystyle\frac{{H_{0}} }{{H}}}} \right) \times
\nonumber \\ \nonumber \\
 &&\times\left[ {{\displaystyle\frac{{A\left( {\rho ,v} \right)}}{{\sin\left( {2\rho}
\right)}}}e^{i\phi} - {\displaystyle\frac{{1}}{{2\rho} }}\left( {1
- {\displaystyle\frac{{i}}{{4\rho} }\frac{{H}}{{H_{0}} }}}
\right)} \right]
\end{eqnarray}
where $H_{0}=m^{2}/e$ is the (2+1)-analogue of the Schwinger
critical  magnetic field introduced initially in the
(3+1)-dimensional electrodynamics, and the functions $\phi $ and
\textit{A($\rho $,v)} are given by the following expressions:
\begin{equation}
\label{eq34}
\phi = \frac{{\omega ^{2}}}{{h}}\left[ {\frac{{\rho
\left( {1 - v^{2}} \right)}}{{2}} - \frac{{\sin\left( {\rho \left(
{1 - v} \right)} \right)\sin\left( {\rho \left( {1 + v} \right)}
\right)}}{{\sin2\rho} }} \right],
\end{equation}
\begin{eqnarray}
\label{eq35}
 A\left( {\rho ,v} \right) \!\!\!\!\!\!\!\!&&= \cos\left( {2\rho v} \right) - i{\displaystyle\frac{{\omega
}}{{m}}}\sin\left( {2\rho v} \right) +
{\displaystyle\frac{{H}}{{4H_{0} \rho} }}\cos\left( {2\rho v}
\right)\left( { - i + {\displaystyle\frac{{\omega ^{2}\rho \left(
{1 - v^{2}} \right)}}{{eH}}}} \right) - \nonumber \\ \nonumber
\\
&& -\left( {{\displaystyle\frac{{\omega} }{{m}}}}
\right)^{2}{\displaystyle\frac{{\sin\left( {\rho \left( {1 + v}
\right)} \right)\sin\left( {\rho \left( {1 - v} \right)}
\right)}}{{\sin^{2}\left( {2\rho}  \right)}}}.
\end{eqnarray}

It is worth mentioning that the above expression for the photon
elastic scattering amplitude takes  both the photon energy
$\omega $ and  the external field intensity $H$ exactly into account,  and
hence, it is valid for the fields of arbitrary intensity.

\subsection{Electron-positron pair photo-production  in an external
magnetic field}

The imaginary and real parts of the  elastic scattering amplitude
in a constant magnetic field (\ref{eq30}) yield,  via the optical
theorem, the rate of electron-positron pair photo-production  and
the photon mass squared, respectively:
\begin{equation}
\label{eq36} w = - 2{\rm Im}(T),
\end{equation}
\begin{equation}
\label{eq37} \delta \left( {m^{2}} \right) = 2\omega {\rm Re}(T)
\end{equation}

Consider now the case of relatively weak magnetic fields and high
energies of photons:
\begin{equation}
\label{eq38}
H \ll H_{0} ,
\quad
m \ll \omega .
\end{equation}

Expanding trigonometric functions in (\ref{eq33})-(\ref{eq35}) in the domain
$\rho \ll 1$, which provides the main contribution to the amplitude in
(\ref{eq33}), we can rewrite the  photon scattering amplitude as follows:
\begin{equation}
\label{eq39}
T = - i\e^{i\pi /4}\frac{{e^{2}m}}{{\left( {4\pi}  \right)^{3/2}\omega
}}\int\limits_{1}^{\infty}  {\frac{{du}}{{u^{3/2}\sqrt {u - 1}} }} \left[ {1
+ \frac{{8u - 5}}{{3}}} \right]\left( {\frac{{\chi} }{{4u}}}
\right)^{1/3}G'\left( {\zeta}  \right),
\end{equation}
where parameter $\chi $ is determined by the relation
\begin{equation}
\label{eq40}
\chi = \frac{{H\omega} }{{H_{0} m}} = \sqrt { - \frac{{e^{2}\left( {F_{\mu
\nu}  k^{\nu} } \right)^{2}}}{{m^{6}}}} .
\end{equation}

We have also introduced the function
\begin{equation}
\label{eq41}
G\left( {\zeta}  \right) = \int\limits_{0}^{\infty}  {\sqrt {y} dy\exp\left(
{ - iy\zeta - iy^{3}/3} \right)} ,
\end{equation}
where $\zeta =(4u/\chi )^{2/3}$. Thus, in the above range of
comparatively weak magnetic fields and high photon energies, the
photon scattering amplitude depends on the external field
strength and photon energy via only single parameter $\chi $. The
function $G(\zeta )$ is a generalization of the well known Airy
function $f\left( {\zeta}  \right) = i\int\limits_{0}^{\infty}
{dx\exp\left( { - iyx - ix^{3}/3} \right)} $ in the amplitude of
elastic photon scattering in the (3+1)-QED in the case $H \ll
H_{0} $, $m \ll k_{ \bot}  $. Now, with the help of Mellin
transformations with the parameter $\lambda = 4\sqrt {3} /\chi $,
we obtain the elastic scattering amplitude:
\begin{equation}
\label{eq42} T\left( {\lambda}  \right) = \frac{{1}}{{2\pi
i}}\int\limits_{\gamma - i\infty} ^{\gamma + i\infty}  {ds\lambda
^{ - s}T\left( {s} \right)},
\end{equation}
\begin{eqnarray}
T\left( {s} \right) \!\!\!\!\!\!\!\!&&= - \frac{{e^{2}m\Gamma
\left( {1/2} \right)}}{{\left( {4\pi}  \right)^{3/2}\omega}
}\frac{{36}}{{5}}\exp\left( { - \frac{{i\pi s}}{{2}}}
\right)\Gamma \left( {1 - \frac{{s}}{{2}}} \right)\Gamma \left(
{\frac{{3s}}{{2}} - \frac{{1}}{{2}}} \right)\times \nonumber \\ \nonumber \\
&&\times\left[ {\frac{{4\Gamma \left( {s} \right)}}{{\Gamma \left(
{s + 1/2} \right)}} - \frac{{\Gamma \left( {s + 1/2}
\right)}}{{\Gamma \left( {s + 3/2} \right)}}} \right]
\label{TAmpl}
\end{eqnarray}
The integral (\ref{eq42}), according to \cite{K28}, can be expressed
in terms of Meijer
$G$-functions \cite{meijer}.

In the framework of the general study of the structure of the
photon polarization operator in (2+1)-dimensional QED in a
constant magnetic field, expressions for the symmetric and
antisymmetric parts of the PO were obtained in \cite{K26,K28}.

In the case of relatively weak fields and high photon energies, when
\begin{equation}
\label{K4}
H\ll H_0=\frac{m^2}{e},\; \omega \gg m,
\end{equation}
the photon elastic scattering amplitude in (2+1)-QED has the following
asymptotics \cite{K28}
\begin{equation}
\label{K5} T=\left\{\begin{array}{l} -i{\displaystyle\frac{e^2 m
\exp\left(-i\frac{\pi}6\right)}{\omega
(4\pi)^{3/2}}}{\displaystyle\frac{24}{5}}
\Gamma\left({\displaystyle\frac{1}{2}}\right)\Gamma\left({\displaystyle\frac{1}{3}}\right)
\left({\displaystyle\frac{\chi}{4\sqrt{3}}}\right)^{1/3},\; \chi \gg1,\\[5pt]
-{\displaystyle\frac{e^2
m}{360\pi\omega}}\chi^2-i{\displaystyle\frac{e^2 m}{8\pi\omega}}
\left({\displaystyle\frac{3\chi\pi}{8}}\right)^{1/2}\exp\left(-{\displaystyle\frac{8}{3\chi}}\right),\;
\chi \ll 1,
\end{array}\right.
\end{equation}

Compare the results of (\ref{K5}) with the analogous results of
(3+1)-QED \cite{K38}. For $\chi\gg1$, the photon elastic
scattering amplitude (both its real and imaginary parts) in
(2+1)-QED is proportional to  $\chi^{1/3},$ while in (3+1)-QED it
grows as $\propto \chi^{2/3}.$ Thus, for $\chi \gg 1$, the
lowering of space-time dimensionality effectively leads to
suppression of the growth of the one-loop contribution to the
scattering amplitude by the factor  $\chi^{1/3}.$ Nevertheless, no
regular correspondence between changing the space-time
dimensionality and the dependence of the amplitude $A$ on
parameter $\chi$ can be detected. In fact, for $\chi \ll 1$, the
imaginary part of $A$ that according to the optical theorem
determines the rate of electron-positron pair production, in
(2+1)-QED is proportional to
$\chi^{1/2}\exp\left(-{\displaystyle\frac{8}{3\chi}}\right),$
while in (3+1)-QED the preexponential factor is $\propto \chi.$ As
for the real part of the amplitude $T,$ it appears to be
proportional to $\chi^2$ for $\chi\gg1$ in both cases.
 The above conclusions demonstrate that the magnetic properties of
 photons change
significantly with the  reduction of the  space-time dimensionality
from (3+1) to (2+1).

\section{ Radiative electron energy shift in (2+1)-QED}

\subsection{ Electron mass shift in (2+1)-QED in external magnetic field
without Chern-Simons term}

The radiative shift of the electron ground state energy in an external
constant magnetic field $H$ in the framework of (2+1)-QED was studied
in \cite{K29}, where the cases with  $\theta=0$ and $\theta\ne0$
(topologically massive QED) were studied separately.

\begin{figure}
\begin{center}
\includegraphics[scale=0.8]{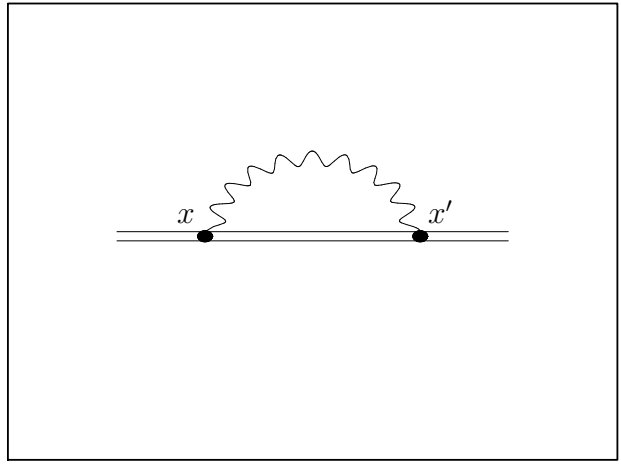}
\begin{picture}(0,0)(0,0)
\put(-170,90){$x$} \put(-75,90){$x'$}
\end{picture}
\caption{\label{SE}The electron ground state energy shift
$\varepsilon _{0}$ in an external magnetic field in the one-loop
approximation.}
\end{center}
\end{figure}

The radiative shift $\Delta \varepsilon _{0}$ of the electron ground
 state energy $\varepsilon _{0}$ in an external magnetic field
in the one loop approximation is depicted in Fig. \ref{SE}. It is
described by the following integral:
\begin{equation}
\label{eq1}
\Delta \varepsilon = \frac{{1}}{{T}}\int {d^{3}xd^{3}x'\bar {\varphi} _{n}
\left( {x} \right)M\left( {x,x'} \right)\varphi _{n} \left( {x'} \right)}
\end{equation}
where
\begin{equation}
\label{eq2_}
M\left( {x,x'} \right) = - ie\gamma ^{\mu} G\left( {x,x'} \right)\gamma
^{\nu} D_{\mu \nu}  \left( {x,x'} \right)
\end{equation}
is the mass operator in the one-loop approximation, $T$ is a sufficiently
long time interval, during which the process proceeds, $\varphi _{n}$ is
the electron
wave function in the external electromagnetic field, $G(x,x')$ and
$D_{\mu \nu}(x,x')$ are the electron and photon propagators in the external
field.

Consider the electron ground state, when $\varepsilon _{0}=m$, where $m$ is the
electron mass, the electron charge $e= - e_{0}<0$.

We will consider for the external field using the exact solutions of the
Dirac equation as the electron wave function and also the exact
propagators of the particles in external field.
For the field given by the potential in the gauge (\ref{eq13}), we employ the electron
propagator in the form (\ref{eq14}). The two-component wave function
of an electron in
the magnetic field is as follows:
\begin{equation}
\label{eq3_}
\varphi _{0} \left( {x} \right) = L^{ - 1/2}\left( {h/\pi}
\right)^{1/4}\exp\left[ { - imt - h\left( {x_{1}}  \right)^{2}/2}
\right]\left( {{\begin{array}{*{20}c}
 {0} \hfill \\
 {1} \hfill \\
\end{array}} } \right),
\end{equation}
where  $L$ is the normalization length.
First, we consider the case without a topological term in the QED
Lagrangian, which is
most close to QED in (3+1) dimensions. Recall that in the (2+1)-QED the
electron spectrum in a magnetic field is purely discreet
(\ref{discr}).  We write the photon propagator in the Feynman gauge:
\begin{equation}
\label{eq4} D_{\mu \nu}  \left( {k} \right) ={\displaystyle\frac{
g_{\mu \nu}} {k^{2} + i0}},
\end{equation}
where $k$ is the  virtual photon momentum. Calculations, similar to
those for the (3+1)-QED case, and subtraction of the divergent term $\Delta
\varepsilon \left| {_{H \to 0}}  \right.$ yield the following integral
representation of the electron energy shift:
\begin{equation}
\label{eq5} \Delta \varepsilon _{0} = \frac{{e^{2}}}{{8\pi
^{3/2}\sqrt {f} }}\int\limits_{0}^{\infty}  {\frac{{dy}}{{\sqrt
{y}} }} \int\limits_{0}^{1} {\frac{{du}}{{\sqrt {u}} }} \exp\left(
{ - \frac{{uy}}{{f} }} \right)\left[ {\frac{{2 - u + 2u\e^{ -
2y}}}{{1 - u +{\displaystyle\frac
        {u\sinh{(y)}}{y}}\e^{ - y}}} -
\left( {2 + y} \right)} \right],
\end{equation}
where the field parameter is $f =e_{0}H/m^{2}$. Now
consider the asymptotic behavior of $\Delta \varepsilon _{0}$ in the case
of strong, $f \gg 1$, and weak, $f \ll 1$, fields.

In the case $\theta=0$, the asymptotic expressions for the
radiative electron  mass shift have the form \cite{K29}:
\begin{equation}
\label{K7}
\bigtriangleup\varepsilon_0=\left\{\begin{array}{l}
{\displaystyle\frac{e^2}{8\pi}}f\left(2+\ln{\bigg(\displaystyle\frac{f}{2}\bigg)}\right),\; f={\displaystyle\frac{eH}{m^2}} \ll 1,\\
{\displaystyle\frac{e^2}{8\pi}}\ln{(2f)},\; f\gg1.
\end{array}\right.
\end{equation}

It is convenient to introduce the so called magnetic susceptibility,
defined as follows:
\begin{equation}
\label{eq8_a}
\chi _{2 + 1} = \frac{{\partial} }{{\partial f} }\Delta \varepsilon
_{0}.
\end{equation}
Then, we obtain it in two limiting cases:
\begin{equation}
\label{eq8} \chi _{2 + 1} = \frac{{\partial} }{{\partial f}
}\Delta \varepsilon _{0} = \frac{{e^{2}}}{{8\pi} }\left\{
{{\begin{array}{*{20}c}
 {\ln{(f)} ,\;f \ll 1}, \hfill \\
 {{\displaystyle\frac{1}{f}} ,\;f \gg 1}. \hfill \\
\end{array}} } \right.
\end{equation}

Comparing the asymptotic behavior of the magnetic susceptibility
$\chi={\displaystyle\frac{\partial \varepsilon_0}{\partial f}}$
with the corresponding results for (3+1)-QED \cite{K38}:
\begin{equation}
\label{eq9}
\chi_{3 + 1} = \frac{{e^{2}m}}{{\left( {4\pi}  \right)^{2}}}\left\{
{{\begin{array}{*{20}c}
 { - 1,\;f \ll 1}, \hfill \\
 {\left( {\displaystyle\frac{2}{f}}\right)\ln{(2f)} ,\;f \gg 1}, \hfill \\
\end{array}} } \right.
\end{equation}
we can observe the following:\\
1) in (2+1)-QED in strong field $(f\gg1)$, it decreases faster,
i.e., $(\propto f^{-1}),$ than in (3+1)-QED
$\left(\propto{\displaystyle\frac{\ln{(f)}}{f}}\right);$\\
2) in the weak field limit in (3+1)-QED the result for $\chi$ is
finite and does not depend on the field in the limit  $f \ll 1,$
while in (2+1)-QED, the effective magnetic susceptibility diverges
at the origin as  $\ln{(f)}.$ This divergence, according to
\cite{K29}, is cancelled by consideration for the Chern--Simons
term in (\ref{K1}).

Thus, lowering the space-time dimensionality leads to considerable
changes in the behavior of the magnetic properties of electrons and to
changes in radiative effects that accompany the photon propagation in
a constant external field.

\subsection{Electron mass shift in topologically massive (2+1)-QED in
external magnetic field}

Now let us turn to the topologically massive QED with the
Chern-Simons term $\theta  \ne 0$ \cite{K29}, making use of the
photon propagator in the Landau gauge (\ref{eq6}). Then, instead
of the topologically trivial case with $\theta =0$ (\ref{eq5}), we
obtain the following expression for the electron mass shift in a
magnetic field in topologically massive (2+1)-QED:
\begin{equation}
\label{eq11_}
{\begin{array}{l}
 \qquad \quad \Delta \varepsilon _{0} ={\displaystyle\frac{e^{2}}{8\pi ^{3/2}\sqrt {f}
}}{\displaystyle\int\limits_{0}^{1} {\frac{{du}}{{\sqrt {u}} }}}
{\displaystyle\int\limits_{0}^{\infty} {\frac{{dt}}{{\sqrt {t}}
}}} \e^{ - ut/f} \Bigg\{{\displaystyle\frac{{1}}{{F}}}\Bigg[
 \e^{ - v}\bigg( {2 - u + 2u\e^{ - 2t}} \bigg) +\\
 {\;} \\
 +{\displaystyle\frac{{\e^{ - v} - 1}}{{v}}}\bar {u}\left(
 {1 - u\bigg( {{\displaystyle\frac{{3}}{{2}}} + {\displaystyle\frac{{\e^{ - 2t}}}{{F}}} + {\displaystyle\frac{{t\left( {2
- u} \right)}}{{f} }}} \bigg) +
 {{\displaystyle\frac{{\theta} }{{m}}}\e^{ - 2t}\bigg( {1 - {\displaystyle\frac{{2ut}}{{f} }} +
{\displaystyle\frac{{2}}{{F}}}}\bigg)}}\right)\Bigg]  - R_{0}\Bigg\},\\
\end{array}}
\end{equation}
where
\begin{equation}
\label{eq12_}
R_{0} = \e^{ - v}\left( {2 + u} \right) + \frac{{\e^{ - v} - 1}}{{v}}\bar
{u}\left[ {1 - u\left( {\frac{{5}}{{2}} + \frac{{t\left( {2 - u}
\right)}}{{f} }} \right) + \frac{{\theta} }{{m}}\left( {3 -
\frac{{2ut}}{{f} }} \right)} \right],
\end{equation}
\begin{equation}
\label{eq13_}
\bar {u} = 1 - u,\;v = \frac{{\theta ^{2}}}{{H}}\frac{{\bar {u}t}}{{u}},\;F
= \bar {u} + u\frac{{\sh t}}{{t}}\e^{ - t}.
\end{equation}

The above expression for the $\Delta \varepsilon _{0}$ accounts for the
external field effect exactly. Note that the $\Delta \varepsilon _{0}$
is the function of two parameters: the field parameter
$f=e_{0}H/m^{2}$ and the mass parameter
\begin{equation}
\label{3.14}
\mu ={\theta \over m}.
\end{equation}
Considering the asymptotic behavior of the electron energy shift
in a weak magnetic field: $f \ll 1$, $\mu \ll 1$, we obtain the
result:
\begin{equation}
\label{eq14_} \Delta \varepsilon _{0} = \frac{{e^{2}}}{{8\pi} }f
\left( {2 + \ln{\bigg({\displaystyle\frac{{f} }{{2}}}\bigg)} +
f\left( {\lambda}  \right)} \right),
\end{equation}
where
\begin{equation}
\label{eq15_} \lambda  = {\mu \over f},\,\, f\left( {\lambda}
\right) = \left( {1 + \lambda}  \right)^{2}\ln\left( {1 + \lambda}
\right) - \lambda - \lambda \left( {\lambda + 2}
\right)\ln{(\lambda)}
\end{equation}
and $\lambda$ has an arbitrary value. When $\lambda =0 (\theta
=0),$ $f(\lambda )$ vanishes
 and (\ref{eq14_}) reduces to (\ref{eq6}), calculated without Chern-Simons
term in
the Lagrangian, i.e. with $ \theta =0$. It is in accordance with the gauge
invariance of the theory (compare (\ref{eq4}) and (\ref{eq6})).

When $\lambda \gg 1$, the energy shift (\ref{eq14_}),
(\ref{eq15_}) writes as follows:
\begin{equation}
\label{3.17} \Delta \varepsilon _{0} = \frac{{e^{2}}}{{8\pi} }f
\bigg( {\ln{(\mu)} + {\rm const}} \bigg),\quad f\ll \mu \ll 1.
\end{equation}
Hence, the magnetic susceptibility $\chi _{2+1}$ in weak magnetic fields,
when $f  \to 0$, becomes finite due to the topological term
$\theta  \ne 0$ that cures logarithmic divergence
of the magnetic susceptibility $\chi _{2+1}$, taking place for $\theta =0$.

\subsection{Electron self-energy in (2+1)-dimensional topologically massive QED at
finite temperature and density}

Here, we will calculate the  contributions of the
finite temperature and density to the electron self energy in (2+1)-QED with
the Chern-Simons term,
represented in the Lagrangian (\ref{K1}). The dependence of  the dispersion
law on the
 finite temperature and matter density
\begin{equation}
\label{eq16}
E^{2} = \vec {p}^{2} + m^{2} + \Delta \left( {E^{2}} \right)
\end{equation}
is represented by the same formulas \cite{K37} as
in (3+1)-QED:
\begin{equation}
\label{eq17_} \Delta \left( {E^{2}} \right) = -
\frac{{1}}{{2}}{\rm Tr}\bigg[ {\left( {\hat {p} + m} \right)\Sigma
\left( {p} \right)} \bigg].
\end{equation}
We use the following expression for the electron mass operator
\begin{equation}
\label{eq18_}
\Sigma \left( {p} \right) = ie^{2}\int {\frac{{d^{3}k}}{{\left( {2\pi}
\right)^{3}}}\gamma ^{\mu} S\left( {p - k} \right)} \gamma ^{\nu} D_{\mu \nu
} \left( {k} \right)
\end{equation}
and perform our calculations in the framework of the the real-time
finite temperature technique. We choose the
temporal electron and  photon Green functions  in the Landau gauge
\cite{K37}
\begin{equation}
\label{eq19_}
D_{\mu \nu}  \left( {k} \right) = - \left[ {\frac{{1}}{{k^{2} - \theta ^{2}
+ i0}} + 2\pi \frac{{\delta \left( {k^{2} - \theta ^{2}}
\right)}}{{\e^{\left( {\left| {k_{0}}  \right|/T} \right)} - 1}}}
\right]\left( {g_{\mu \nu}  - \frac{{k_{\mu}  k_{\nu} } }{{k^{2} + i0}} +
i\theta \varepsilon _{\mu \nu \lambda}  \frac{{k^{\lambda} }}{{k^{2} + i0}}}
\right),
\end{equation}
\begin{equation}
\label{eq2_0}
S\left( {p} \right) = i\frac{{\hat {p} + m}}{{p^{2} - m^{2} + i\varepsilon
}} - 2\pi \delta \left( {p^{2} - m} \right)\left( {\hat {p} + m}
\right)\left[ {\frac{{\theta \left( {p_{0}}  \right)}}{{\e^{\left( {p_{0} -
\mu}  \right)/T} + 1}} + \frac{{\theta \left( { - p_{0}}  \right)}}{{\e^{ -
\left( {p_{0} - \mu}  \right)/T} + 1}}} \right],
\end{equation}
with the following notations  adopted in this Section: $T$ is the
temperature and $\mu $ is the chemical potential. With the use of
the above formulas we obtain the
following result for the contribution of the effects of
finite temperature and
density  to the real part of the fermion energy shift:
\begin{eqnarray}
\label{eq2_1}
\Delta \left( {E^{2}}\right) =
2e^{2}{\displaystyle\int
\frac{d^{3}k}{\left({2\pi}\right)^{3}}}\bigg[ {m^{2} + m\theta +
{\displaystyle\frac{{pk}}{{k^{2}}}}\left(
{k^{2}-pk}\right)}\bigg]\times \nonumber \\ \nonumber \\
 \times\left\{ {{\displaystyle\frac{{\delta \left( {k^{2} - 2pk} \right)}}{{k^{2} - \theta
^{2}}}}\left[ {{\displaystyle\frac{{\theta \left( {q_{0}}
\right)}}{{\e^{\left( {q_{0} - \mu}  \right)/T} + 1}}} +
{\displaystyle\frac{{\theta \left( { - q_{0}} \right)}}{{\e^{ -
\left( { - q_{0} - \mu}  \right)/T} + 1}}}}\right] -
{\displaystyle\frac{{\delta \left( {k^{2} - \theta ^{2}}
\right)}}{{k^{2} - 2pk}}}{\displaystyle\frac{{1}}{{\e^{\left|
{k_{0}} \right|/T} - 1}}}} \right\},
\end{eqnarray}
where $q_\mu =p_\mu-k_\mu $. The term with the step-function $\theta
(q_{0})$ accounts for the contribution of the electron gas  and
the term with $\theta (q_{0})$ represents the contribution of
the positron gas. Upon integration with the help of
$\delta-$functions, leaving only the electron contribution, we obtain:
\begin{equation}
\label{eq2_2} \Delta \left( {E^{2}} \right) =
\frac{{e^{2}}}{{8\pi} }\int\limits_{}^{} {\frac{{d\varepsilon}
}{{\e^{\left( {\varepsilon - \mu} \right)/T} + 1}}\left( {1 -
\frac{{2m^{2} + 2m\theta + \theta ^{2}/2}}{{\sqrt {\left( {p_{0}
\varepsilon + \theta ^{2}/2 - m^{2}} \right)^{2} - \left( {pq}
\right)^{2}}} }} \right)},
\end{equation}
where $\varepsilon = \sqrt {\vec {q}^{2} + m^{2}} $. Hence, the following
expression for the completely degenerate electron gas is obtained:
\begin{equation}
\label{eq2_3} \Delta \left( {E^{2}} \right) =
\frac{{me^{2}}}{{8\pi} }\left[ {\left( {\frac{{\mu} }{{m}} - 1}
\right) - 2\left( {1 + \Theta}  \right)^{2}\ln\left| {\frac{{\sqrt
{\left( {{\displaystyle\frac{{\mu} }{{m}}}} \right)^{2} +
b{\displaystyle\frac{{\mu} }{{m}}} + c} +
{\displaystyle\frac{{\mu} }{{m}}} +
{\displaystyle\frac{{1}}{{2}}}b}}{{\sqrt {1 + b + c} + 1 +
{\displaystyle\frac{{1}}{{2}}}b}}} \right|} \right],
\end{equation}
where
\begin{equation}
\label{eq2_4}
\Theta = \frac{{\theta} }{{2m}},
\quad
E_{p} = \sqrt {1 + \frac{{\vec {p}^{2}}}{{m^{2}}}} ,
\quad
b = 2E_{p} \left( {2\Theta ^{2} - 1} \right),
\quad
c = E_{p} ^{2} + 4\Theta ^{4} - 4\Theta ^{2}.
\end{equation}

In massless (2+1)-QED, when $m \to 0$, with the finite Chern-Simons term, the
dispersion law is as follows
\begin{equation}
\label{eq2_5}
E^{2} = p^{2} + \frac{{e^{2}}}{{8\pi} }\left[ {\mu - \frac{{\theta} }{{2\pi
}}\left( {\sqrt {4p\mu + \theta ^{2}} - \theta}  \right)} \right].
\end{equation}

In order to calculate the effective electron  mass in the charge symmetric
case at finite temperature, we should sum up series of the following form:
\begin{equation}
\label{eq2_6}
S\left( {q,y} \right) = \sum\limits_{n = 1}^{\infty}  {q^{n}\e^{ny}{\rm{Ei}}\left[ {
- ny} \right]} ,
\end{equation}
where $\left| {q} \right| < 1$, Ei[$-x$] is the integral exponential
function and the parameter $y$ is proportional to the inverse
temperature $T^{-1}$. Introducing the polilogarithm function
Li$_{s}[q]$, we obtain with the help of the  method developed in
\cite{vsh}, \cite{K37}
\begin{equation}
\label{eq2_7} S\left( {q,y} \right) = \int\limits_{c - i\infty}
^{c + i\infty} {\frac{{ds}}{{2\pi i}}\frac{{\pi \Gamma \left( {s}
\right)}}{{\sin{(\pi s)}}}y^{ - s}{\rm {Li}}\left[ {q} \right]} ,
\quad 0 \ll c \ll 1
\end{equation}
where
\begin{equation}
\label{eq2_8}
{\rm {Li}}_{s} \left[ {q} \right] = q\Phi \left[ {q,s,1} \right] = \sum\limits_{n =
1}^{\infty}  {\frac{{q^{n}}}{{n^{s}}}} .
\end{equation}

Then closing the integration contour to the right (with $y \gg 1$) or to the
left (with $y \ll1$),  we obtain the representation of
$S(q,y)$ in the form of rapidly converging  series of residues of the
integrand at the points $s=1,2,3,\dots$ and $s=0,-1,-2,-3,\dots$
respectively. Thus for the mass shift of the electron at rest we finally
obtain
\begin{equation}
\label{eq2_9} {\begin{array}{c} \Delta \left( {m^{2}} \right) =
{\displaystyle\frac{{e^{2}}}{{4\pi} }}\left\{
 { - T\,{\rm {Li}}_{1} \left[ { - \e^{ - m/T}} \right] +{\displaystyle\frac{{\left( {\theta + 2m}
\right)^{2}}}{{4m}}}\bigg( {S\left[ { - \e^{ - m/T},y_{1}} \right]
- S\left[
{ - \e^{ - m/T},y_{2}}  \right]} \bigg) +} \right. \\
{\;} \\
+\left. {T\,{\rm {Li}}_{1} \left[ {\e^{ - \theta /T}} \right] -
{\displaystyle\frac{{\left( {\theta + 2m}
\right)^{2}}}{{4m}}}\bigg( {S\left[ {\e^{ - \theta /T},y_{3}}
\right] - S\left[ {\e^{ - \theta /T},y_{4}}  \right]} \bigg)}
\right\},
 \end{array}}
\end{equation}
where  parameters $y_{i}$ $ (i=1,2,3,4)$ are defined as
\begin{equation}
\label{eq3_0}
y_{1} = \frac{{\theta ^{2}}}{{2mT}},
\quad
y_{2} = \frac{{1}}{{T}}\left( {2m - \frac{{\theta ^{2}}}{{2m}}} \right),
\quad
y_{3} = \frac{{1}}{{T}}\left( {\theta + \frac{{\theta ^{2}}}{{2m}}} \right),
\quad
y_{4} = \frac{{1}}{{T}}\left( {\theta - \frac{{\theta ^{2}}}{{2m}}}
\right).
\end{equation}

Let us discuss now two interesting results obtained in
\cite{K36,K37,K35}, where the  finite temperature and density
contributions to the radiative shift of the electron energy in the
topologically massive (2+1)-QED were calculated.

Expressions (\ref{eq2_8}), (\ref{eq2_9}) yield the principal logarithmic
approximation for
the electron mass shift in the limiting case of high temperature.
In the charge symmetric case, the one-loop radiative shift of the
electron mass has the following asymptotic behavior  \cite{K36,K37}:
\begin{equation}
\label{K8} \delta m =\left\{\begin{array}{l}
{\displaystyle\frac{e^2}{8\pi}}\left(\ln{\bigg({\displaystyle\frac{\theta}{2m}}\bigg)}
+{\displaystyle\frac{T}{m}}\left(\ln{(2)}-1\right)-
{\displaystyle\frac{T}{m}}\ln{\left({\displaystyle\frac{\theta}{T}}\right)}\right),\; \theta \ll 2m \ll T,\\
{\displaystyle\frac{e^2}{2\pi}}\left({\displaystyle\frac{T}{\theta}}\ln{\left(1-\e^{-\theta/T}\right)}-
{\displaystyle\frac{T}{\theta}}\ln{\left(1+\e^{-m/T}\right)}\right),\;
2m \ll \theta, \; 2mT \ll \theta^2.
\end{array}\right.
\end{equation}

This result at low temperatures $(T \ll
{\displaystyle\frac{\theta^2}{2m}},\; 2m \ll \theta)$ admits a
limiting transition to massless electrodynamics \cite{K36,K37}:
\begin{equation}
\label{K9} \delta m \bigr|_{m
\to0}=\frac{e^2}{2\pi}\frac{T}{\theta}\left[\ln\left(1-
\e^{-\theta/T}\right)-\ln{(2)} \right].
\end{equation}

It follows from (\ref{K9}) that, at finite temperature in the
initially massless (2+1)-QED with the Chern--Simons term, there
exists an effect of the fermion mass generation in virtue of the
radiative effects.

As follows from  (\ref{K8}), (\ref{K9}), the temperature mass
shift at low temperatures is negative. However, the total
radiative electron mass shift also includes a part which is
independent of temperature and is determined by the Chern-Simons
term \cite{K3},\cite{K37}:
\begin{equation}
\label{eq3_4}
\Delta m_{\rm {total}} = \Delta m\left( {T = 0} \right) + \Delta m\left( {T \ne 0}
\right) = {{e^{2}}\over {2\pi}}  + \Delta m\left( {T} \right).
\end{equation}

Under the conditions (\ref{K8}), we have
\begin{equation}
\label{eq3_5}
\left| {\frac{{\Delta m\left( {T} \right)}}{{\Delta m\left( {T = 0}
\right)}}} \right| \propto \frac{{T}}{{\theta} } \ll 1,
\quad
T \ll \theta ,
\end{equation}
i.e. the temperature correction to the electron mass is small as compared to the
$\Delta m (T=0)$ at $T \ll \theta $ and the total radiative
electron mass shift remains positive. In the case of high temperature
(\ref{K8}),  we
obtain qualitatively different behavior of the temperature mass shift:
\begin{equation}
\label{eq3_6} \left| {\frac{{\Delta m\left( {T} \right)}}{{\Delta
m\left( {T = 0} \right)}}} \right| \propto
\frac{{T}}{{m}}\ln{\bigg(\frac{{T}}{{\theta} }\bigg)} \gg 1, \quad
T \gg 2m \gg \theta .
\end{equation}

Thus, with growing temperature, the mass shift $\Delta m(T)$ changes its sign
and its contribution to the radiative electron mass shift  becomes
dominant.

Finally, let us come back to the dispersion law and calculate the exchange
correction to the thermodynamic potential:
\begin{equation}
\label{eq3_7} \frac{{\Omega _{\rm{exch}}} }{{V}} = \int
{\frac{{d^3p}}{{\left( {2\pi} \right)^{3}}}\frac{{\Delta \left(
{E^{2}} \right)}}{{2\sqrt {\vec {p}^{2} + m^{2}}}
}}\bigg(\e^{\left( {\frac{{E_{p} - \mu} }{{T}}} \right)} +
1\bigg)^{-1} .
\end{equation}

In the case with $m=0$, $T=0$, i.e. massless (2+1)-QED at zero
temperature, the dispersion law is given by (\ref{eq2_5}) and the integration
immediately results in
\begin{equation}
\label{eq3_8} \frac{{\Omega _{\rm {exch}}} }{{V}} =
\frac{{e^{2}}}{{32\pi ^{2}}}\left\{ {\mu ^{2} - \frac{{\theta}
}{{2}}\left[ {2\left( {\sqrt {4\mu ^{2} + \theta ^{2}} - \theta}
\right) + \theta \ln{\bigg(\frac{{\sqrt {4\mu ^{2} + \theta ^{2}}
- \theta} }{{\sqrt {4\mu ^{2} + \theta ^{2}} + \theta} }\bigg)} -
2\theta \ln{\bigg(\frac{{\mu} }{{\theta} }\bigg)}} \right]}
\right\}.
\end{equation}

It should be noted that whereas the one-loop mass operator in (2+1)-QED
without the Chern-Simons term is infrared divergent on the mass shell, so that
charge screening effects should be taken into account
\cite{novikov,K37}, the mass operator in topologically massive QED at
finite
temperature and density is finite already in the one-loop approximation.
Moreover, when $\theta  \to 0$, the dispersion law (\ref{eq2_5}) becomes
\begin{equation}
\label{eq3_9}
E^{2} = p^{2} + \frac{{e^{2}\mu} }{{8\pi} } \quad ,
\end{equation}
i.e. the finite density effects produce a gap in the spectrum. However, at
any finite value of $\theta $, the dispersion law looks as
\begin{equation}
\label{eq40_}
E^{2}\left( {p,\theta \ne 0} \right)\buildrel {p \to 0} \over
\longrightarrow 0
\end{equation}
and the gap is not produced at least in the one-loop
approximation. Thus, with the mass parameter  $\theta$ tending to
zero, there appears a gap in the electron energy spectrum that is
due to finite density effects at zero temperature. However any
finite value of this parameter prevents formation of a gap in the
energy spectrum \cite{K37}.

\section{Induced parity-violating thermal effective action}

 As it was pointed out in the Introduction, a fundamental property of
 the gauge field action in odd-dimensional space-times is its
 non-invariance under
``large'' (homotopically non-trivial) gauge transformations.
Hence, the expression $\exp{(iS_{cs})}$ will have a unique value
only if a quantization condition is imposed on the Chern--Simons
coefficient (the topological mass). This condition leads to
quantization of the coefficient in units of $g^2/(4 \pi)$, which
results in restoration of invariance of the theory as a whole
under large gauge transformation. The above conclusions are valid
only for zero temperature quantum field theory. The situation
seriously changes when temperature becomes finite. This is due to
the fact that, at finite temperature, the fermion radiative
corrections shift the tree level CS coefficient and become
unquantized  continuous functions of temperature \cite{2,bra,cab}.
Hence, at first glance, the problem of large gauge symmetry
restoration seems to have no solution. However, in a number of
recent publications \cite{D2,D3,D4,5}, it was demonstrated that,
although the temperature dependent Chern--Simons coefficient may
violate the invariance, there exist the other terms in the action
which compensate this violation, so that the total effective
action restores its gauge invariance. It turned out that 
the effective action can be constructed in such a way that it
 becomes invariant under both small
and large gauge transformations at any temperature. This becomes
 possible by the use of a suitable
regularization of the Dirac operator determinant, at the price of
 emerging parity anomalies.

As for the perturbative approach, it appears impossible to evaluate the
effective action in a closed form, contrary to the (0+1) model, and
this can be done only for a restricted class of backgrounds, for
example, for static fields, when $A_0=A_0(t)$, $\vec A=\vec A(\vec
x)$ \cite{6}.
In the static limit, (${\bf p}\rightarrow 0,$ $p_0=0$), with all
energies vanishing, at any order of perturbation, the ``large''
gauge invariance is not manifest. However, it is possible to sum
up the leading order terms (this can be done through derivation of
the Ward identity for the special case) in the parity violating
effective action in this limit and the resulting effective action
appears to have a form, which exactly coincides with the result
obtained in the frame of special background gauge mentioned above.
Moreover, this action is a generalization of the (0+1)-dimensional
result and is invariant under large gauge transformations
\cite{6}. Therefore, the significance of calculation of the effective
action in 
the special background is that it represents the leading term in
the effective action in the static limit.

In \cite{razum}, with the use of the technique
developed in \cite{4}, an exact expression for this one-loop effective
action was obtained, when the background field configuration was
taken as a superposition of an abelian and a non-abelian gauge
fields in the group $U(1)\times SU(2)$.  Let us discuss the results of
\cite{razum} in more detail.

\subsection{ Problem statement}

The parity breaking part of the action can be written in the form
\begin {equation}
\Gamma_{odd}(A,M)=\frac{1}{2}\bigg(\Gamma(A,M)-\Gamma(A,-M)\bigg),
\label{1}
\end{equation}
where the effective action $\Gamma(A,M)$ is related to the action
for massive fermions $S_F(A,M)$ in the standard way
\[
\exp{\bigg(-\Gamma(A,M)\bigg)}=\int D\psi
D\overline{\psi}\exp{\bigg(-S_F(A,M)\bigg)}.
\]
The background field is the following combination of constant
abelian ($U(1)$) and non-abelian  ($SU(2)$) fields:
$A_{\mu}=gA^{(1)a}_{\mu}T_a+eA^{(2)}_{\mu}I$. Here $T_a$~are the
$SU(2)$ group generators, and~$I$~is the unit matrix in the color
space. Moreover, the following algebra of $\gamma$-matrices is used
$$
\gamma^3=\sigma^3,\ \ \gamma^1=i \sigma^1,\ \ \gamma^2=i \sigma^2,\ \
\gamma^{\mu}\gamma^{\nu}=g^{\mu\nu}-i\epsilon^{\mu\nu\alpha}\gamma_{\alpha}.
$$
Index 3 is related to the Euclidean time coordinate $\tau$.
Expression (\ref{1}) can not be calculated exactly for the case of
arbitrary fields $A$, and hence, we shall restrict ourselves to
solving  the problem in the above mentioned special gauge
\begin {equation}\ba {l}
A^{(1),a}_3 =|A^{(1)}_3(\tau)|n^a,\ \ A^{(2)}_3=A^{(2)}_3(\tau),\\
\nonumber \\
 A^{(1)}_j=A^{(1)}_j({\bf x}),\ \
A^{(2)}_j=A^{(2)}_j({\bf x}) \quad (j=1,2).
\ea\label{3}\end{equation} Here, $n^a$~is a fixed unit vector in
the color space ($n^an^a=1$), and we also require that
\begin{equation}
[A_j,A_3]=0, \quad [A_j,n]=0.
\label{3_}
\end{equation}
Then, the effective action for massive fermions in the background
gauge field (\ref{3}) in the (2+1)-dimensional space-time at
finite temperature can be written in the form
\begin {equation}
S_F(A,M)=\il^{\beta}_0d\tau\int d^2x\,\overline{\psi}(\gamma\partial+i\gamma
A+M)\psi,
 \label{4}
\end{equation}
where $\gamma\partial=\gamma^{\mu}\partial_{\mu}$, $\gamma
A=\gamma^{\mu}A_{\mu}$, and $\beta=1/T$. Fermion and gauge fields
that appear in (\ref{4}) satisfy anti-periodic and periodic
conditions, respectively
$$\ba{c}
 \psi(\beta,{\bf x})=-\psi(0,{\bf x}), \ \
 \overline{\psi}(\beta,{\bf x})=-\overline{\psi}(0,{\bf x}),\\ \\
 A_{\mu}(\beta,{\bf x})=A_{\mu}(0,{\bf x}).
\ea$$

\subsection{ Parity breaking action}

The fermion determinant can be written in the following form
\begin{equation}\ba{c}
{\rm det} (\gamma\partial+i\gamma A+M)={\displaystyle\int D\psi
D\overline{\psi}}\exp\bigg(-{\displaystyle\int\limits^{\beta}_0d\tau}{\displaystyle\int
 d^2x}\overline{\psi}(\gamma\partial+i\gamma A+M)\psi\bigg).
\ea\label{5}\end{equation} Then, we perform the gauge
transformation of fermion fields
\begin{eqnarray}
\ba{l} \psi(\tau,{\bf x}) =
\exp{\bigg(-i\bigg[g\Omega^{(1)}(\tau)n+e\Omega^{(2)}(\tau)I\bigg]\bigg)}\psi'(\tau,{\bf
x}),\nonumber \\ \nonumber \\
\overline{\psi}(\tau,{\bf x}) = \overline{\psi}\,'(\tau,{\bf
x})\exp{\bigg(i\bigg[g\Omega^{(1)}(\tau)n+e\Omega^{(2)}(\tau)I\bigg]\bigg)}.
\ea
\end{eqnarray}
Since only the third components $A_3$ depend on $\tau$ in the
gauge field configuration in question, and $A_j$ are independent
of $\tau$, and also by virtue of condition~(\ref{3_}), the
transformations of this kind do not act on spatial components of
the potential, and time dependence of the third component can be
excluded, if we choose $\Omega^{(1)}$ and $\Omega^{(2)}$ in the
form
$$\ba{l}
\Omega^{(1)}(\tau)=-{\displaystyle\il^{\tau}_0}
d\tau'A^{(1,n)}_3(\tau')+\bigg({\displaystyle\frac{1}{\beta}}{\displaystyle\il^{\beta}_0}d\tau'
A^{(1,n)}_3(\tau')\bigg)\tau,\\
\Omega^{(2)}(\tau)=-{\displaystyle\il^{\tau}_0}
d\tau'A^{(2)}_3(\tau')+\bigg({\displaystyle\frac{1}{\beta}}{\displaystyle\il^{\beta}_0}d\tau'A^
{(2)}_3(\tau')\bigg)\tau.
\end{array}$$
Thus, the fermion determinant (\ref{5}) takes the form
\begin {equation}
{\rm det}(\gamma\partial+i\gamma A+M)=\int D\psi
D\overline{\psi}\exp{\bigg(-S_F(A_j,\tilde{A}_3,M)\bigg)},
\label{6}\end{equation} where the action is equal to
\begin{equation}\ba{l}
S_F(A_j,\tilde{A}_3,M)= \ds=\int\limits^{\beta}_0 d\tau\int
d^2x\,\overline{\psi}\bigg(\gamma\partial+i(\gamma_jA_j+\gamma_3\tilde{A}_3)+M\bigg)
\psi, \ea\label{7}\end{equation} and the quantity
$\tilde{A}_3={\displaystyle\frac{1}{\beta}}{\displaystyle\int\limits^{\beta}_0}d\tau\bigg(gA^{(1)}_3(\tau)n+eA^{(2)}_3(\tau)I\bigg)
$ takes a constant value now. To calculate the determinant
(\ref{6})-- (\ref{7}), perform Fourier-expansion of the fermion
fields
\begin{eqnarray}
\psi(\tau,{\bf x})=\fr{1}{\beta}\sum^{n=+\infty}_{n=-
\infty}e^{i\omega_n\tau}\psi_n({\bf x}),\nonumber \\ \nonumber \\
\overline{\psi}(\tau,{\bf
x})=\fr{1}{\beta}\sum^{n=+\infty}_{n=-\infty}
e^{-i\omega_n\tau}\overline{\psi}_n({\bf x})
\end{eqnarray}
and then write the action as
\[\ba{l}
S_F(A_j,\tilde{A}_3,M)==\fr{1}{\beta}\sum^{n=+\infty}_{n=-\infty}\!\int\!\!
d^2x\,\overline{\psi}({\bf x})_n \bigg(\gamma
d+i\gamma_3(\omega_n{+}\tilde{A}_3)+M\bigg)\psi_n({\bf x}). \ea\]
Here, we introduced the following notation for the differentiation
operator $\gamma d=\gamma_j(\partial_j+iA_j)$, and $\omega_n=\pi
(2n+1)/\beta$~are Matsubara frequencies for fermions. Thus, taking
into account that the fermion measure can now be written as
$$
D\psi(\tau,x)D\overline{\psi}(\tau,x)
=\prod^{n=+\infty}_{n=-\infty}D\psi_n({\bf x})D\overline{\psi}_n({\bf x}),
$$
we obtain for the fermion determinant
\[
{\rm det}(\gamma\partial+i\gamma A+M)
=\prod^{n=+\infty}_{n=-\infty}{\rm det}\bigg(\gamma
d+M+i\gamma_3(\omega_n+\tilde{A}_3)\bigg),
\]
where the determinant behind the product sign can be written in the
form
\begin {equation}
\int D\chi_n D \overline{\chi}_n \exp\bigg(\!-\!\!\int\!
d^2x\,\overline{\chi}_n({\bf x})(\gamma d+\rho_n
e^{i\gamma_3\phi_n})\chi_n({\bf x})\bigg).
\label{11}\end{equation} Here, we made use of the Euler formula
with the notations $\rho_n=\sqrt{M^2+(\omega_n{+}\tilde{A}_3)^2}$
and
$\phi_n=\arctan\bigg({\displaystyle\frac{\omega_n+\tilde{A}_3}{M}}\bigg)$.
To calculate determinant (\ref{11}), the known method of
calculating the anomalous Fujikawa Jacobian ~\cite{8} can be used.
To this end, we first make the following transformation of spinors
$\chi$ (chiral rotation in the space $(1+1)$)
\[
\chi_n(x)=
\exp{\bigg(-i{\displaystyle\frac{\phi_n}{2}}\gamma_3\bigg)}\chi'_n(x),\,\,
 \overline{\chi}_n(x)= \overline{\chi}'_n(x)\exp{\bigg(i{\displaystyle\frac{\phi_n}{2}}\gamma_3\bigg)}.
\]
Then, it is easily verified that expression (\ref{11}) takes the form
\begin {equation}
{\rm det}\bigg(\gamma
d+M+i\gamma_3(\omega_n+\tilde{A}_3)\bigg)=J_n {\rm det}(\gamma
d+\rho_n), \label{13}\end{equation} where
\[
J_n=\exp\lt\{-\frac{i}{4\pi}{\rm tr}\left(\phi_n\int d^2x
\epsilon_{ij}\Big(F^{(1)}_{ij}+\frac{1}{2}F^{(2)}_{ij}\Big)\right)\rt\}.
\]
In this expression, $F^{(1)}$ and $F^{(2)}$ are the non-abelian and
abelian field tensors, respectively. The last factor in (\ref{13})
does not depend explicitly on the sign of the fermion mass and
hence does not contribute to the parity breaking part of the
effective action. Therefore, the expression for $\Gamma_{odd}$ can
be written right away
\begin {equation}
\Gamma_{odd}=\frac{i}{4\pi}\tr\left(\Big(\sum^{n=+\infty}_{n=-\infty}\phi_n\Big)
\int d^2x \epsilon_{ij}\Big(F^{(1)}_{ij}+\frac{1}{2}F^{(2)}_{ij}\Big)\right).
 \label{15}
\end{equation}
Moreover, since the field $\phi_n$ itself should be expanded in color
 space directions
 $\phi_n=\phi^0_nI+\phi^a_nT_a$, then with its explicit expression
 known, expression for its every  color component can be easily
 obtained
\[
\phi^0_n=\frac{1}{2}\arctan\left({\displaystyle\frac{2M(\omega_n+e\tilde{A}^{
(2)}_3)} {M^2+{\displaystyle\frac{g^2}{4}}|\tilde{A}^{(1)}_3|^2-
(\omega_n+e\tilde{A}^{(2)}_3)^2}}\right),
\] \[
\phi^a_n=\arctan\left({\displaystyle\frac{gM|\tilde{A}^{(1)}_3|}
{M^2-{\displaystyle\frac{g^2}{4}}|\tilde{A}^{(1)}_3|^2+
(\omega_n+e\tilde{A}^{(2)}_3)^2}}\right)n^a.
\]
Here, it should be emphasized that, for the combination of an abelian and
 non-abelian fields in question,  unlike the case of a pure
 non-abelian field, not only the color component of the field
 $\phi^a_n$ but also component $\phi^0_n$ will contribute to
 $\Gamma_{odd}$.  This is related to the fact that the abelian field
 tensor $F^{(2)}$ is present in the integrand of the expression
 (\ref{15}).  In other words, the expression behind $\tr$ in
 (\ref{15}) can be rewritten as
\begin {equation}
\int d^2x
\epsilon_{ij}\left(\bigg(\sum^{n=+\infty}_{n=-\infty}\phi^a_n\bigg)F^{(1)a}_{ij}
+\bigg(\sum^{n=+\infty}_{n=-\infty}\phi^0_n\bigg)\frac{1}{2}F^{(2)}_{ij}\right).
\label{18}
\end{equation}
Now, consider calculation of one of the sums in  (\ref{18}), for
 instance $\sum\phi^a_n$, in more detail. To this end, introduce the
 following notations  $m=\beta M$,
 $x={\displaystyle\frac{g}{2}}\beta|\tilde{A}^{(1)}_3|$ and
 $y=e\beta |\tilde{A}^{(2)}_3|$. Then,
\[
\sum^{n=+\infty}_{n=-\infty}\phi^a_n =\sum^{n=+\infty}_{n=-\infty}
\!\arctan\lt(\!\frac{2mx}{m^2-x^2+\bigg((2n+1)\pi+y\bigg)^2}\!\!\rt).
\]
Next, we use the equivalent form of this expression~\cite{4}
\[
\sum(x,y,m)=\il^{x}_0 du\frac{\partial\Sigma}{\partial u}(u,y,m),
\]
where
$$
\fr{\partial\Sigma}{\partial u}(u,y,m)=2m\sum^{n=+\infty}_{n=-
\infty}\frac{m^2+u^2+((2n+1)\pi+y)^2}{[m^2+((2n+1)\pi+y)^2-u^2]^2+4m^2u^2}.
$$
This expression is exactly equal to
\begin {equation}\ba{l}
\fr{\partial\Sigma}{\partial u}(u,y,m)= \ds{}=-\frac{m}{2\pi i
}\oint\limits_C dz\tanh{\Big(\frac{z}{2}\Big)}
{\displaystyle\frac{m^2+u^2+(y-iz)^2}{[m^2+(y-iz)^2-u^2]^2+4m^2u^2}},
\ea\label{22}\end{equation} where contour $C$ encloses all the
poles of $\tanh\lt({\displaystyle\frac{z}{2}}\rt)$, i.~e. points
$z=i(2n+1)\pi$. Next, contour $C$ should be replaced by the
equivalent sum $C_1+C_2$ of two other contours, which will
together enclose only four singular points of the fraction. Thus,
summing up residues at all these points, we obtain the required
expression~(\ref{22})
\begin{equation}
\fr{\partial\Sigma}{\partial
u}(u,y,m)=\frac{\sinh(m)}{4}\left[\frac{1}{\cosh(m)+\cos(u-y)}
+\frac{1}{\cosh(m)+\cos(u+y)}\right]
\end{equation}
and so, upon integration over $u$~\cite{9}, the expression for the
sum $\sum\phi^a_n$ takes the form
\begin{equation}
\ds\sum^{n=+\infty}_{n=-\infty}\phi^a_n=\arctan\left(\tanh\Big(\fr
m2\Big)\tan\Big(\fr{x-y}2\Big)\right)+\arctan\left(\tanh\Big(\fr
m2\Big)\tan\Big(\fr{x+y}2\Big)\right).
\end{equation}
It is clear that the sum  $\sum\phi^0_n$ can be calculated in a
similar way. Combining them, the final form of the expression
 ~(\ref{15}) can be written as follows
\begin{equation}
\Gamma_{\rm odd}=\Gamma^{(1)}+\Gamma^{(2)}, \label{25}
\end{equation}
where \begin{eqnarray} &&\Gamma^{(1)} =
\fr{ig}{8\pi}\bigg(I_1+I_2\bigg)n^a \int d^2\,x
\epsilon_{ij}F^{(1)a}_{ij},\label{26} \\ \nonumber \\
&&\Gamma^{(2)} = \fr{ie}{8\pi}\bigg(I_1-I_2\bigg)\int d^2x\,
\epsilon_{ij}F^{(2)}_{ij} \label{27}
\end{eqnarray}
and
$$
I_{1,2}=\arctan{\left(\tanh\Big(\fr{\beta M}{2}\Big)
 \tan\Big({\displaystyle\frac{g\beta}{4}}|\tilde{A}^{(1)}_3|
 \pm
 {\displaystyle\frac{e\beta}{2}}|\tilde{A}^{(2)}_3|\Big)\!\right)}.
$$

It is easily seen that, in the limiting case, when one of the fields
(either abelian or non-abelian) vanishes, the obtained expression
exactly renders the earlier results~\cite{4}. One can also easily
observe that, in the zero temperature limit, the parity breaking
part of the action (\ref{25}), (\ref{26}), (\ref{27}) goes over
into the half-sum of the Chern-Simons terms for abelian and
non-abelian fields, each of which reproduces the known results of
the zero temperature quantum field theory~\cite{10,niemi}
\[
\Gamma_{\rm
odd}|_{T=0}=\frac{1}{2}\frac{M}{|M|}\bigg(S_{CS}^{(1)}+S_{CS}^{(2)}\bigg),
\]
where in our case
$$\ba{l}
S_{CS}^{(1)}=\fr{ig^2}{4\pi}\tr\int
d^3x\,A_3^{(1)}\epsilon_{ij}F_{ij}^{(1)},\\[9pt]
S_{CS}^{(2)}=\fr{ie^2}{4\pi}\int d^3x\,A_3^{(2)}\epsilon_{ij}F_{ij}^{(2)}.
\ea$$

\section{Conclusions}

In the present paper, we reviewed some results of investigations of
radiative effects in the (2+1)-dimensional quantum
electrodynamics and Yang-Mills theory with consideration for the
influence of external
fields, finite temperature and matter density.

In particular, a possibility of the Chern-Simons term generation
due to non-zero external field as well as finite temperature and
density in the (2+1)-dimensional QED was demonstrated. The radiative
shift of the photon topological mass  in (2+1)-QED in an external
magnetic field  as a function of the field strength and photon
energy was analyzed. The calculation and analysis of the rate of
the electron-positron pair photoproduction in (2+1)-QED in an
external magnetic field was demonstrated and the probability of
the process 
was presented. The electron mass shift in (2+1)-QED in an
external magnetic field was calculated as a function of the
electron energy and the fields strength. The Chern-Simons term
contribution was also investigated. The calculation of the
effective magnetic sensitivity of the 2D electron gas was
performed, and the Chern-Simons term was shown to eliminate its
divergence in weak external fields. The gap in the electron
spectrum in (2+1)-QED was shown to appear due to the finite
electron gas density effects. The topological term was shown to
eliminate the gap and make the radiative mass shift converge in
(2+1)-QED at finite temperature.

With the aim of obtaining the above results, detailed calculation
of the photon polarization operator in (2+1)-QED in external
magnetic field was performed. In particular, the case of relatively weak
magnetic fields and high photon energies,  $H \ll H_{0} $, $m \ll
\omega $, was investigated. Comparison of the results, obtained
for the photon elastic scattering in (2+1)-QED with the
corresponding results in (3+1)-QED, demonstrated that the increase
of the photon elastic scattering amplitude with growing parameter
$\chi = H\omega /\left( {H_{0} m} \right)$, when $\chi \gg 1$, in
(2+1)-QED is determined by the factor $\chi ^{1/3}$, whereas in
(3+1)-QED it increases as $\chi ^{2/3}$.  In the opposite case,
$\chi \ll 1$, the imaginary part of the scattering amplitude, responsible
for the e$^{+}$e$^{ - }$ pair photo-production, behaves as $\chi
^{1/2}\exp{(-{\displaystyle\frac{8}{3\chi}})}$, whereas in
(3+1)-QED, the corresponding preexponential is equal to $\chi $.
However, in the case $\chi \ll 1$, the real part of the photon
scattering amplitude, which determines the photon mass $\Delta
\left( {m^{2}} \right) = 2\omega {\rm Re}(T)$,   is proportional
to $\chi ^{2}$, and this coincides with the result of (3+1)-QED.

Calculation of the one-loop radiative energy shift of the ground
state of an electron in a constant magnetic field in (2+1)-QED
with the topological term was presented and the asymptotic behavior of
the effective magnetic sensitivity $\chi _{2 + 1} = \partial
\left( {\Delta \varepsilon _{0}}  \right)/\partial \beta $ of the
2D electron gas was studied. Comparison of the obtained results
for the magnetic sensitivity in the special case of topologically
massless (2+1)-QED, $\theta=0 $, with those in (3+1)-QED reveals
the following changes in the behavior of the magnetic
susceptibility $\chi_{2+1} $ as a function of the field parameter
$\beta =e_{0}H/m^{2}$. In relatively weak fields, $\chi _{2+1}$
diverges logarithmically for $\beta  \to   0$, whereas $\chi
_{3+1}$ tends to a constant. In strong fields, $\chi _{2+1}$
decreases with growing field intensity faster than the
corresponding function $\chi_{3+1}$ does. The consideration for
the topological term $\theta  \ne 0$ eliminates logarithmic
divergence of the magnetic sensitivity $\chi _{2+1}$ in weak
magnetic fields ($\beta \to 0 $), i.e., $\chi _{2 + 1} \propto
\ln\left( {\theta /m} \right)$. At finite temperature and density,
consideration for the Chern-Simons term makes the electron mass
operator finite already in the one-loop approximation, whereas it
is infrared divergent on the mass shell in the one-loop approximation
in the topologically massless (2+1)-QED.

We also presented the results of considering the influence of  finite
temperature and density on the radiative
electron energy shift in the (2+1)-dimensional topologically massive
QED. The case of a
completely degenerate electron gas was studied. In the charge
symmetric case, the electron effective
mass at finite temperature was calculated and
the limiting cases of low $T \ll \theta ^{2}/\left( {2m} \right)$ and high $\theta \ll
2m \ll T$
temperatures were investigated. The  mass
shift due to finite temperature, $\Delta m(T)$, was shown to be
negative at low temperatures. However, it is
small as compared to the zero temperature term $\Delta m(T=0)$,
determined by the photon topological mass, so the total radiative
electron mass shift remains positive. With growing temperature, the
term $\Delta m(T)$ changes its sign and increases, so its contribution
to the radiative electron  mass
shift  becomes prevailing. Moreover,
when $\theta  \to 0$, the dispersion law becomes
$E^{2} = p^{2} + e^{2}\mu /\left(
{8\pi}  \right)$, i.e. the finite density effects produce a gap in the
electron
spectrum. However, at any finite value of $\theta $, the energy
squared $E^{2}\left( {p,\theta
\ne 0} \right)\left| {_{p \to 0} \to}  \right.0$ and, at least in the
one-loop approximation, the gap is no more present.

In addition to the above results, the problem of possible breaking
 of gauge invariance  of the effective action due to
the one-loop corrections to the CS coefficient in the
(2+1)-dimensional (non-)abelian gauge theory at finite temperature
was also considered in the present paper.
For this purpose, the exact expression for the parity breaking
part of the finite temperature (2+1)D effective action 
 induced by massive fermions in the
particular background, consisting of both abelian and non-abelian
gauge fields, has been obtained. The special background
configuration is characterized by the vanishing electric field and
the time-independent magnetic field. In the limiting cases, when
the abelian or non-abelian gauge field vanishes, the obtained general formula
goes over into the earlier results obtained specifically for a non-abelian or
an abelian fields respectively~\cite{4}. Likewise, in the 
limit of vanishing temperature,  we received
the  zero temperature result obtained in earlier
 publications\cite{10,niemi}. It
should be noticed, that our result explicitely demonstrates
only the invariance  under small gauge transformations that do not mix
spatial and time components of the background fields. The
large gauge invariance can be restored in the same way, as it was done
 in \cite{D1}--\cite{D4}, i.e., through consideration of
the full effective action, not restricted by the 1-loop approximation.

Thus, the following general conclusions can be made. The magnetic
properties of photons change significantly with the reduction of
the space-time dimensionality from (3+1) to (2+1). The radiative
electron mass shift in a magnetic field and the effective magnetic
sensitivity in the (2+1)-QED demonstrate that magnetic properties
of electrons also change substantially, when the dimensionality is
reduced from (3+1) to (2+1). Consideration for the finite
topological photon mass $\theta  \ne 0$ eliminates divergences in
the electron mass operator and effective magnetic sensitivity,
whereas finite temperature and density  lead to new physical
effects, such as appearance of the gap in the  electron spectrum.

One of the general conclusions is also that large gauge invariance of
the parity violating effective action at finite temperature in the
(2+1)-dimensional space-time can not be obtained in the framework of the
perturbative approach. This  can only be achieved through  
consideration for contributions of all terms of all orders  in 
the abelian or non-abelian gauge field  action.

\section*{Acknowledgments}

We would like to thank Professors D. Ebert and M. Mueller-Preussker
for many valuable comments and suggestions, and also for their hospitality
at the HU-Berlin.  One of the authors (A. R.) acknowledges
financial support by the Leonhard Euler program
of the German Academic Exchange Service (DAAD) extended to him, while
part of this work was carried out.
The other author (V.Ch.Zh.) acknowledges support by DAAD and
partly by the DFG-Graduiertenkolleg ``Standard Model''. This work was
also supported in part by
the DFG project 436RUS 113/477.

\end{document}